\documentclass{article}
\usepackage{spconf,amsmath}
\usepackage{amssymb}
\usepackage{todonotes}
\usepackage{tabulary}
\usepackage{tabularx}
\usepackage{multirow}
\usepackage{stfloats}
\usepackage{gensymb}
\usepackage{blindtext, graphicx}
\usepackage{multirow}
\usepackage{tikz,pgfplots,pgfplotstable}
\usepackage[bookmarks=false]{hyperref}
\usepackage{textgreek}
\pgfplotsset{compat=1.5,width=10cm}
\usepgfplotslibrary{external}
\usepackage{verbatim}
\usepackage[binary-units=true]{siunitx}
\colorlet{A1}{red!80!white}
\colorlet{B1}{orange!80!white}
\colorlet{C1}{blue!80!white}
\colorlet{A2}{red!80!black}
\colorlet{B2}{orange!80!black}
\colorlet{C2}{blue!80!black}

\usepackage{xargs}  

\usepackage{lipsum}
\usepackage{ifthen}
\usepackage{caption}
\usepackage{pdftexcmds}
\usepackage{stringstrings}
\usepackage{multicol}
\usepackage{graphicx}
\usepackage{booktabs}

\usepackage{algorithm2e}

\usepackage{tabularx}
\usepackage{amsfonts}

\usepackage{enumitem}
\usepackage{xspace}

\definecolor{bblue}{HTML}{4F81BD}
\definecolor{rred}{HTML}{C0504D}
\definecolor{ggreen}{HTML}{9BBB59}
\definecolor{ppurple}{HTML}{9F4C7C}

\newcommand{\scatgan}{ScatGAN\xspace}

\newcommand{\cmmnt}[1]{-}

\newcommandx{\info}[2][1=]{\todo[linecolor=green,backgroundcolor=green!25,bordercolor=green,#1]{#2}}
\newcommandx{\unsure}[2][1=]{\todo[linecolor=red,backgroundcolor=red!25,bordercolor=red,#1]{#2}}
\newcommandx{\change}[2][1=]{\todo[linecolor=blue,backgroundcolor=blue!25,bordercolor=blue,#1]{#2}}
\newcommandx{\improvement}[2][1=]{\todo[linecolor=Plum,backgroundcolor=Plum!25,bordercolor=Plum,#1]{#2}}
\newcommand{\hide}[1]{}

\title{ScatGAN for Reconstruction of Ultrasound Scatterers\\ using Generative Adversarial Networks}

\name{Andrawes Al Bahou, Christine Tanner, Orcun Goksel\thanks{Funding provided by Innosuisse and Swiss National Science Foundation.}}
\address{Computer-assisted Applications in Medicine, Computer Vision Lab, ETH Zurich, Switzerland}

\begin{document}
\maketitle
\begin{abstract}
Computational simulation of ultrasound (US) echography is essential for training sonographers.
Realistic simulation of US interaction with microscopic tissue structures is often modeled by a tissue representation in the form of point \emph{scatterers}, convolved with a spatially varying point spread function. 
This yields a realistic US B-mode speckle texture, given that a scatterer representation for a particular tissue type is readily available. This is often not the case and scatterers are nontrivial to determine.
In this work we propose to estimate scatterer maps from sample US B-mode images of a tissue, by formulating this inverse mapping problem as \emph{image translation}, where we learn the mapping with Generative Adversarial Networks using a US simulation software for training. 
We demonstrate robust reconstruction results, invariant to US viewing and imaging settings such as imaging direction and center frequency. 
Our method is shown to generalize beyond the trained imaging settings,
demonstrated on \textit{in vivo} US data. 
Our inference runs orders of magnitude faster than optimization-based techniques, enabling future extensions for reconstructing 3D B-mode volumes with only linear computational complexity. 
\end{abstract}

\section{Introduction}

Ultrasound (US) is one of the most widely adopted medical imaging technologies for its nonionizing, affordable, portable, and real-time nature. However, US imaging involves challenges with probe navigation and image interpretation, thus requires comprehensive training. 
Traditional training with synthetic phantoms is often unrealistic; and training with cadavers or patients involves ethical issues. 
In contrast, model-based numerical US simulation can allow for training in virtual reality, enabling exposure to complex medical scenarios and rare pathologies.
Thus, developing US echography simulators which can produce images indistinguishable from real ones in real-time has been a major research interest.

Wave-based methods~\cite{USsimulation} for US simulation model complex full-wave propagation, which is not suitable for real-time image synthesis. Convolution-based approaches~\cite{field2} model the entire pulse transmit and echo beamforming pipeline with a spatially-varying Point-Spread Function (PSF), the convolution of which, with a particular \emph{tissue representation}, then yields the desired image.
Ray-based methods~\cite{gpu-raytracing} can, in addition, compute the incident acoustic power by tracing the complex reflections and refractions which the wavefront undergoes, during large-scale interactions with geometrical boundaries of anatomical structures.

US speckle texture is the interference pattern resulting from echos scattered by uncountably many sub-wavelength tissue structures, such as cell nuclei, organelles and large proteins, often referred to as ``scatterers''. Faithful simulation of US texture is crucial, as it carries tissue-specific descriptive and diagnostic information. 
For a realistic texture appearance, however, a viable tissue (scatterer) representation is necessitated.
This paper focuses on extracting such scatterer maps from given image examples.
\begin{figure}[t]
    \includegraphics[height=.33\columnwidth]{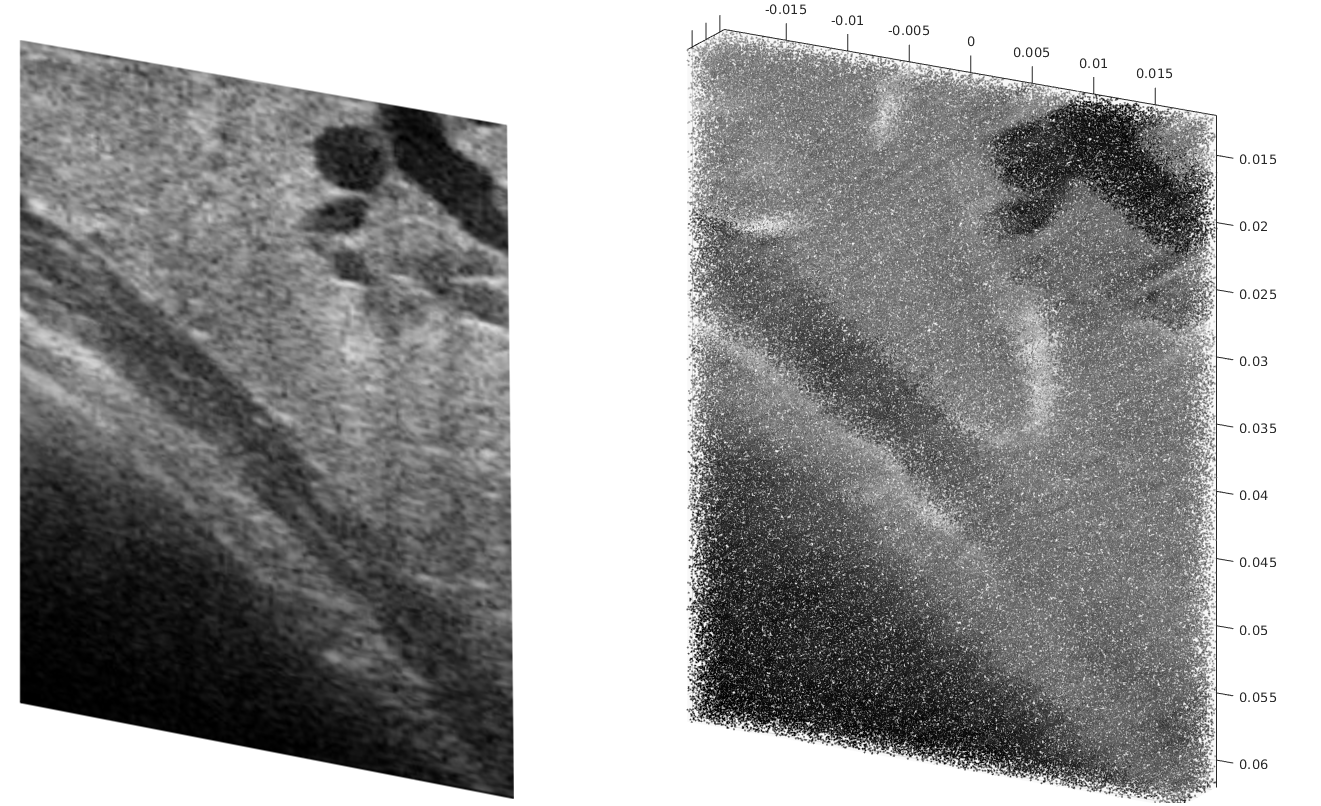}
    \centering
    \caption{Inverse problem of US simulation: Reconstruction of point scatterers (right) from \textit{in vivo} B-mode image (left)}
    \label{fig:front_page}
\end{figure} 
Although a scatterer representation can be defined as a generative model parametrized by a statistical (e.g. Gaussian) distribution~\cite{tissue_parametrization}, arbitrary tissue textures especially with structural information would not necessarily follow well-defined parametric models.
Even if this were valid, e.g. for simple homogeneous tissue patches, 
finding a realistic parameter instantiation by trial-and-error would be tedious. 
An estimated scatterer map that we aim herein would help for an automatic setting. 
Accordingly, in this work we propose to construct a tissue scatterer model given a B-mode image of the region of interest, as illustrated in Fig.~\ref{fig:front_page}. 
In a different context, this is also known as ultrasound deconvolution~\cite{taxt} and it involves the solution of an \textit{inverse} problem~\cite{mattausch}.
State-of-the-art methods for this problem use tedious optimization approaches; which, being exponential in problem size, is not extendable to large images or three dimensions.

We propose a novel pipeline by formulating the inverse problem of US simulation as an image translation task.
To that end, we adopt a learning approach based on Conditional Generative Adversarial Networks \cite{pix2pix}. 
Due to the lack of ground-truth scatterer maps, we employ a synthetically generated training dataset.
We utilize CT images to generate artificial scatterer maps, as the Hounsfield units of CT images relate to tissue density and hence acoustic impedance that cause perturbations and scattering during US echo propagation, and capture a broad range of anatomical visual semantics. 
To the best of our knowledge, this is the first work to tackle the inverse problem of US simulation with a learning-based approach.

\section{Background}
\label{sec:related_work}
\noindent{\bf Forward Ultrasound Simulation.}
\label{sub:forward_us_simulation}
US scattering response from subwavelength particles can be simulated via convolution with a PSF.
A set of infinitesimally small points (\textit{scatterers}) in space can be used as an abstract tissue representation (\textit{scatter map}).
Each scatterer has an amplitude in (0,1], defining the proportion of incident acoustic power that it scatters back. We call this scatter map $g(\mathbf{x})$,  where $\mathbf{x}$$=$$[x,y,z]$ are the lateral, elevational, and axial coordinates. 
A spatially-varying PSF $h(\mathbf{x})$ can be estimated from transducer and imaging parameters, and then convolved with $g(\mathbf{x})$ to obtain a modulated radio-frequency (RF) signal
\begin{equation}
\label{eq:convolution}
    f([x,0,z]) = g([x,y,z]) \ast h([x,0,z]) + \gamma([x,0,z])
\end{equation}
which indicates an image plane at $y$=$0$ resulting from convolution with a 3D PSF, distorted by noise $\gamma$.
Field\,II~\cite{field2} toolbox allows for computing such PSFs.
To generate a grayscale B-mode image $b$, demodulation (envelope extraction) and dynamic-range (logarithmic) compression are applied to $f$. 

\vspace{1ex}
\noindent{\bf Inverse Problem of Scatterer Reconstruction.}
Finding a scatter map $g$ given an RF image can be seen in discrete space as the inverse problem of deconvolution~\cite{taxt,mattausch}, i.e.
$\mathbf{H}\mathbf{g}+\mathbf{\Gamma}$$=$$\mathbf{f}$,
where $\mathbf{H}$ is the convolution matrix induced by the PSF, $\mathbf{g}$ is a vector of scatterer amplitudes, $\mathbf{f}$ is the resulting RF image, and $\mathbf{\Gamma}$ is the noise vector. 
Given the much higher resolution of the scatterer space w.r.t. the image discretization, the solution becomes under-determined, i.e.\ a solution $g$ may not faithfully represent the tissue under different imaging settings, such as viewing directions. 
Mattausch \textit{et al.}~\cite{mattausch} proposed multiple acquisitions for the same field-of-view via electronic beam steering, to obtain multiple equations: $[\mathbf{H}_1, \mathbf{H}_2, \cdots]^T\mathbf{g}$$=$$[\mathbf{f}_1, \mathbf{f}_2, \cdots]^T$, for a better constrained problem:
\begin{equation}
\hat{\mathbf{g}} = arg\min_{\mathbf{g}} ||\mathbf{H}\mathbf{g}-\mathbf{f}||_{1} + \lambda||\mathbf{g}||_1 ~~\mbox{s.t.}~\mathbf{g}\geq 0
\end{equation}
where $\lambda$ is a regularization parameter.
This is solved using the Alternating Direction Method of Multiplies, which yields a robust reconstruction of sparse scatterers on a discrete grid. 

\begin{figure}[t]
    \includegraphics[scale=0.35]{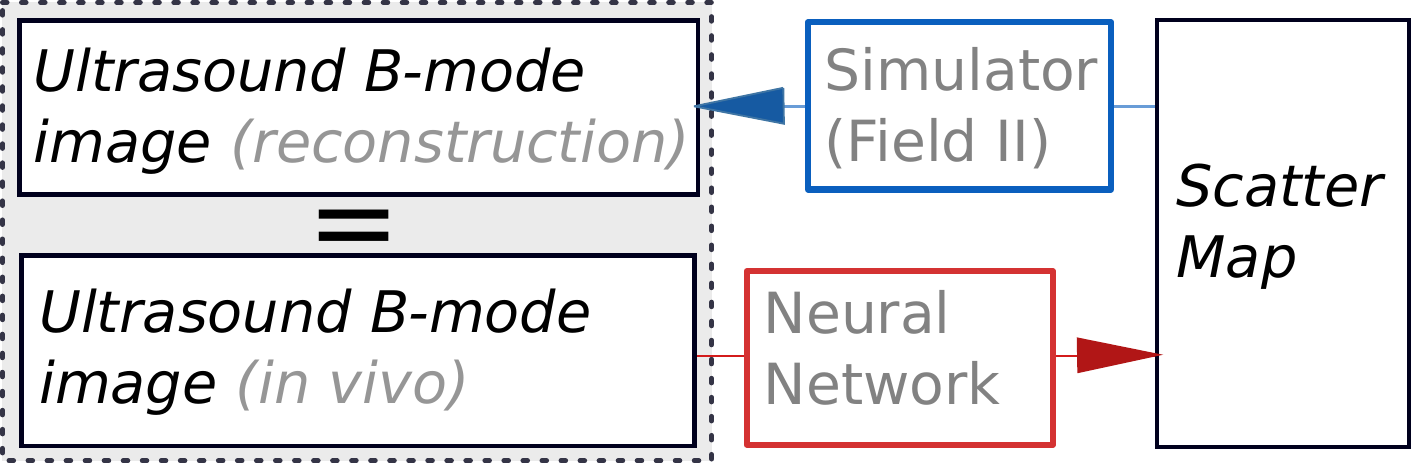}
    \centering
    \caption{\textbf{Forward} problem in blue, \textbf{inverse} problem  in red}
    \label{fig:pipeline}
\end{figure}

\section{Methods}\label{sec:implementation}
We use a learning-based approach to directly map from any B-mode image $b$ to its estimated scatter map $g$. 
Assuming a discrete scatter-map representation, 
we adopt the conditional GAN approach \textbf{pix2pix}~\cite{pix2pix}, which learns an image-to-image translation map, assuming the input and output images are spatially aligned.
CGANs learn a mapping $G(\alpha,\rho):$ $\{\alpha,\rho\} \mapsto \beta$, where $\rho$ is a random input $\rho$$\sim$$ P_{\rho}(\rho)$, $\alpha$ is an input image used to condition $G$, and $\beta$ is the output image.

Training the pix2pix GAN requires a dataset consisting of \{B-mode image, scatter map\} pairs.
Since scatterer maps are abstract constructs, and no ground-truth maps are known for phantom, let alone in-vivo images, we resort to creating our own paired data via simulations using  Field\,II~\cite{field2}. The challenge lies in creating a training set rich in visual features for the algorithm to learn from, without over-fitting to implicit statistical distributions induced by the simulation process in FieldII,  to facilitate generalization to \textit{in vivo} data. 

\vspace{1ex}
\noindent{\bf Training Set Creation.}
$N$ scatterers were instantiated in space uniformly at random.
Given the physical simulation domain size, $N$ is set for a scatterer density of 20\,$mm^{-1}$, 
in order to ensure fully-developed speckle for the given imaging center frequency and settings.
We assign each scatterer $i$ a random amplitude $a_i$ sampled from a normal distribution controlled by an image
$I(\mathbf{x})$$\in$$[0,1]$ via
$\mu(\mathbf{x})$$=$$I(\mathbf{x})$ and
\begin{equation}
\sigma(\mathbf{x}) = -\left|\frac{\sigma_{max}-\sigma_{min}}{2}\left(I(\mathbf{x})-\frac{1}{2}\right)\right|+\sigma_{max}.
\label{eq:sigma}
\end{equation}
The $\sigma(\mathbf{x})$ function is controlled via $\sigma_\mathrm{min}$ and $\sigma_\mathrm{max}$ such that the sampled value $a_i$$\in$[0,1] with a probability of 90\%
for all values $I(\mathbf{x})$. 
Values outside this range are clipped to [0,1], which slightly distorts the normal distribution of amplitudes. Linear interpolation is used to map discrete image pixel locations to continuous scatterer positions, as input to Field\,II.

Initially, we had created $I(\mathbf{x})$
procedurally as additive or multiplicative combinations of various primitive shapes of different intensities. 
However, this process proved insufficient for the learning algorithm to generalize well to \textit{in vivo} images. 
Based on this preliminary observation, we eventually used a set of in-vivo CT images, out of which $I(\mathbf{x})$ is randomly drawn.
This then allowed to naturally present a wide variety of realistic visual features and relate directly to the acoustic impedance that modulate US propagation in tissue. 

\vspace{1ex}\noindent{\bf Network Training.}
\begin{figure}
    \includegraphics[width=0.4\textwidth]{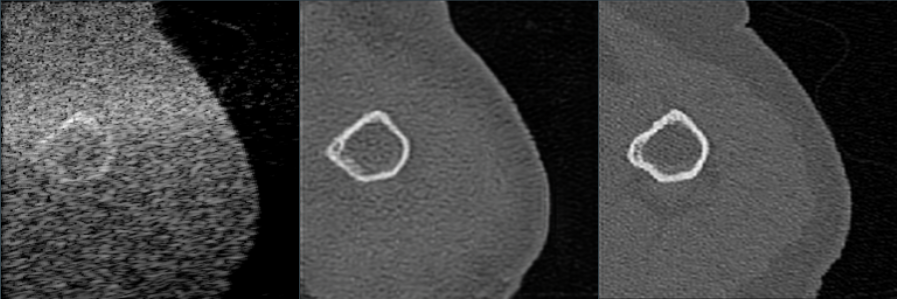}
    \includegraphics[width=0.4\textwidth]{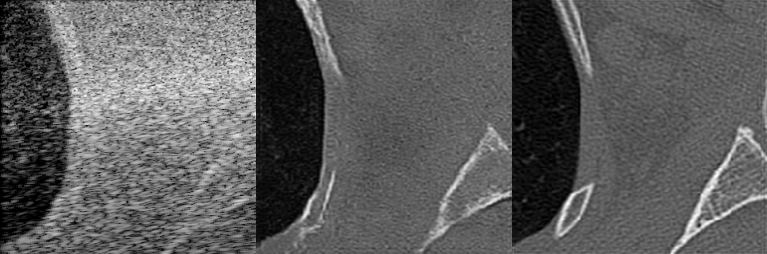}
    \centering
    \caption{Left$\rightarrow$right: B-mode image from training set, scatterer image from ScatGAN1 network, target scatterer image.}
    \label{fig:network_training}
\end{figure}
Our network architecture and optimization follow the original \emph{pix2pix} approach~\cite{pix2pix}.
A 7-layer U-Net
was used as the generator, and a 4-layer patchGAN~\cite{pix2pix} 
as the discriminator. 
The generator was trained to minimize the L1 loss between real and generated scatter images, and the discriminator to maximize the classification accuracy between real and generated scatter images. 
Optimization alternates between discriminator and generator.
Two networks were trained, namely 
``ScatGAN1'' and ``ScatGAN3''. The training set is identical for both and consists of B-mode images of tissue regions imaged at $-10^{\circ}$, $0^{\circ}$ and $10^\circ$.
ScatGAN1 has only a single B-mode image as input, while ScatGAN3 has a concatenation of three B-mode images of the same tissue (at the 3 angles).
Example results from ScatGAN1 are shown in Fig.~\ref{fig:network_training}.

\section{Experiments and Results} 
\label{sec:results}
The ultimate goal of a scatterer map is a realistic B-mode image to be simulated from it.
Therefore, for evaluations we follow a scheme similar to~\cite{mattausch}: We study B-mode image reconstruction errors and similarity by comparing any given B-mode image with its corresponding simulated version from scatterers predicted by \scatgan.
For this, given a B-mode image $b_\mathrm{GT}$, \scatgan network inference is run to obtain a scatter map estimation $\hat{g}$.
This discrete image is then converted to a scatterer point cloud, as described above for Field\,II scatterer preparation. 
Subsequently, a B-mode image $\hat{b}$ is simulated using Field\,II and compared with $b_\mathrm{GT}$ using various similarity metrics.
We use 3 image metrics $F$: Signal-to-noise ratio SNR$=$$\mu/\sigma$, mean image intensity MII$=$$1/|\Omega|\sum_\mathbf{x\in\Omega}b(\mathbf{x})$
and contrast-to-noise ratio CNR$=$$|\mu_{b}-\mu_{d}|/(\sigma_{b}+\sigma_{d})$, where $b$ and $d$ denote selected bright and dark regions. These metrics are compared between $\hat{b}$ and $b_{GT}$, providing normalized errors (incompatibility) 
defined as $100 |{\hat{F}/F_{GT}-1}|$. 
We also compare the histograms (50 bins) of $\hat{b}$ and $b_{GT}$ using the $\chi^2$ distance. 

\begin{figure}[t]
    \includegraphics[trim = 0mm 50mm 0mm 0mm, clip, width=0.33\textwidth]{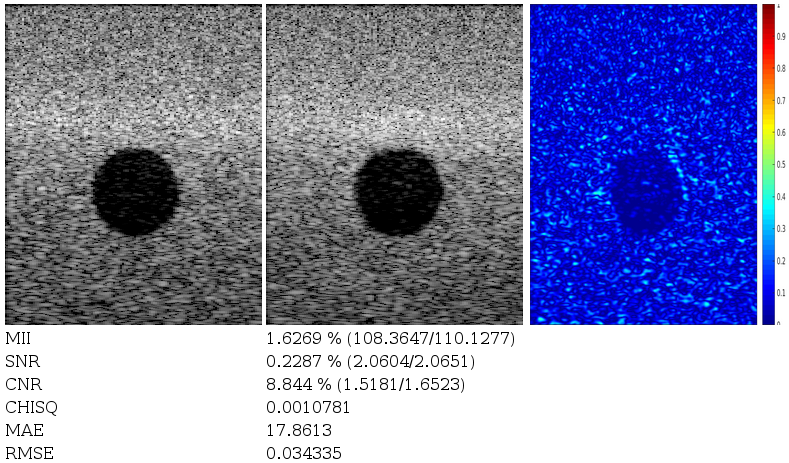}
    \includegraphics[trim = 0mm 50mm 0mm 0mm, clip, width=0.33\textwidth]{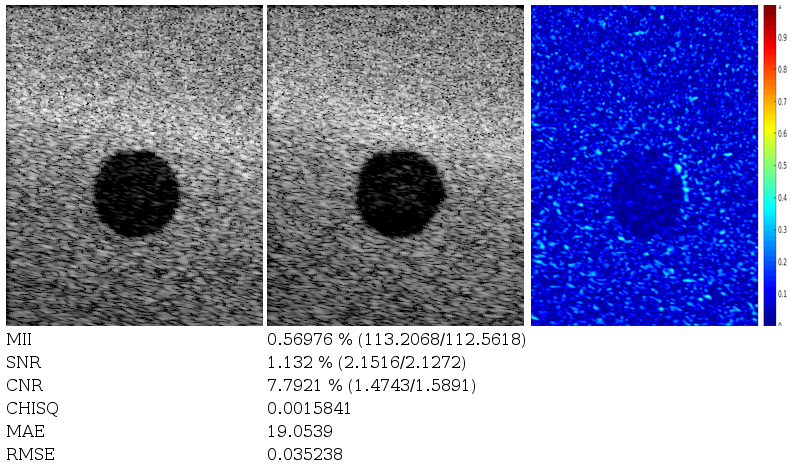}
    \centering
    \caption{Left$\rightarrow$right: Real and reconstructed B-mode image of \textit{in silico} phantom for probe angle of (top) $0^{\circ}$ and (bottom) $10^\circ$ and their normalized absolute error (blue 0\%, red 100\%)}
    \label{fig:inclusion_examples}
\end{figure}

\begin{figure}[tb]
    \includegraphics[trim = 0mm 66mm 0mm 0mm, clip, width=\columnwidth]{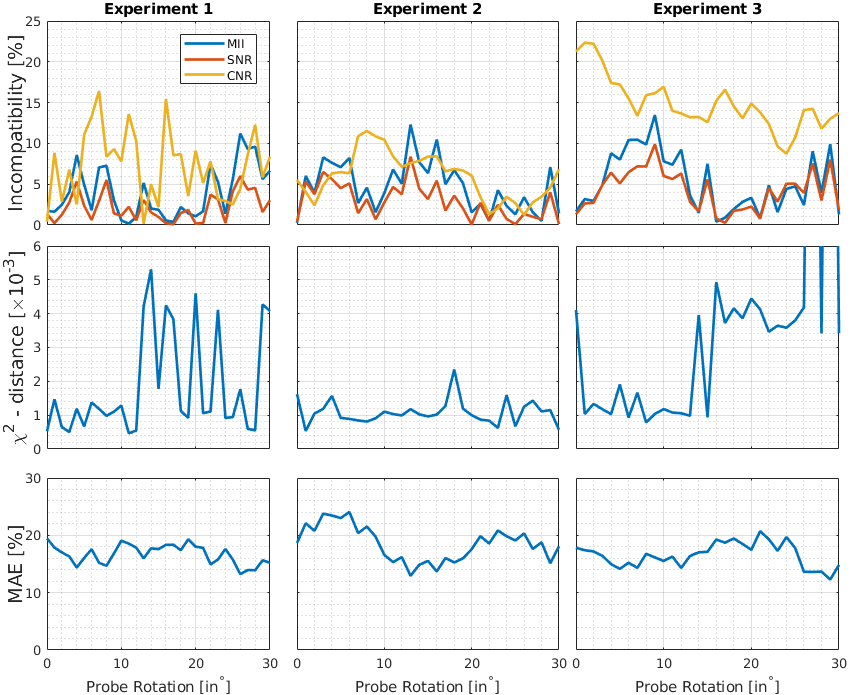}
    \includegraphics[trim = 0mm 0mm 0mm 172mm, clip, width=\columnwidth]{figures/all_cyst_exp.png}
    \centering
    \caption{Experiment results for \textit{in silico} dataset}
    \label{fig:all_cyst_exp}
\end{figure}

\vspace{1ex}\noindent{\bf Numerical experiments.}
For the \textit{in silico} tests, we created a numerical phantom with a circular inclusion of 7\,mm radius, 
with an average scatterer intensity of 10\% of that of the background. 
We simulated RF data in Field\,II for a linear array imaging at 5\,MHz center frequency.
We rotated the probe around the sample to collect 31 unique B-mode images from different viewing directions at angles $\in$ $\{0^\circ, 1^\circ, ..., 30^\circ\}$, see Fig.~\ref{fig:inclusion_examples}.
In {\bf Experiment 1}, we run ScatGAN1 \emph{separately} on each of the 31 B-mode images, to obtain 31 scatter maps.  We then simulate each of these at its corresponding US probe angle, and compare the 31 simulated images with the 31 original B-mode images.
Note that this is simply to test our upper reconstruction bound, and is almost an ``inverse crime'' (although Field\,II and our \scatgan reconstructions are not necessarily inverse of each). 
Fig.~\ref{fig:all_cyst_exp}(left) shows no correlation of error metrics with probe angle, indicating that the reconstruction error is invariant across different images of a similar field-of-view.
In {\bf Experiment 2}, we run ScatGAN1 on the B-mode image simulated at probe angle $0^\circ$, to obtain a single scatter map.
Then, we simulate B-mode images at all 31 angles from this single map. 
This is a realistic setting, where a single scatterer representation will be derived for a tissue region to simulate arbitrary viewing angles.
Note that changing the scatterers at different angles as in Experiment1 creates flickering and discontinuous temporal sequences due to temporal incoherence of independent reconstructions.
Error metrics remain relatively similar across the range of probe rotations in Fig.~\ref{fig:all_cyst_exp}(middle), indicating that the reconstruction is robust for simulating images at other angles.

Multiple viewing angles were found to bring robustness with optimization based methods~\cite{mattausch}.
In \textbf{Experiment 3}, we use ScatGAN3 network (three-image input), where the output is a single scatterer map that explain these three views together. 
This scatter map is then used to simulate all 31 B-mode images ($0^{\circ}$-$30^\circ$). 
ScatGAN3 did not provide extra robustness over ScatGAN1, see Fig.~\ref{fig:all_cyst_exp}(right) vs.\ (middle). 
This is likely due to the capacity of network being reached or more training set being needed with this setting.

\begin{figure}[tb]
    \includegraphics[width=\columnwidth]{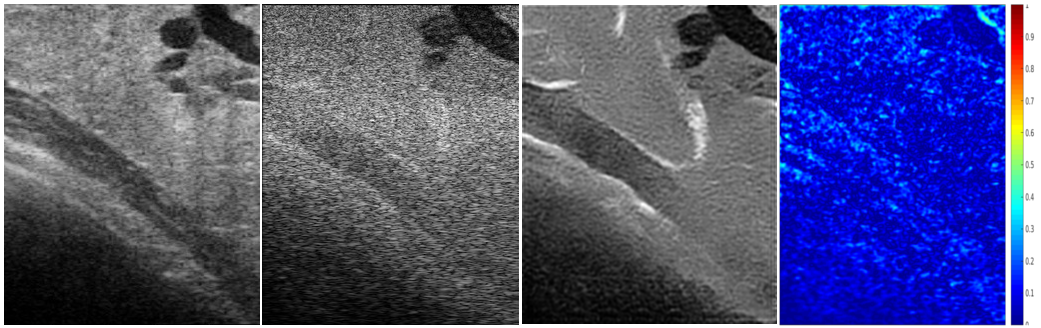}
    \centering
    \caption{Result for \textit{in vivo} image. Left$\rightarrow$right: Input image, reconstructed B-mode image, scatterer image produced by ScatGAN1, normalized absolute error (blue~0\%, red~100\%).}
    \label{fig:in_vivo_3}
\end{figure}

\noindent{\bf In-vivo image results.}
In \textbf{Experiment 4}, we run the network on a real \textit{in vivo} B-mode image.  From this scatterer map, a B-mode image was simulated at the original center frequency of 10~MHz, with successful reconstructions seen in Fig. \ref{fig:in_vivo_3}.


\begin{table}
\setlength{\tabcolsep}{4pt}
\centering
\caption{Comparison of \scatgan with iterative optimization based deconvolution method (ScatRec1,7) from~\cite{mattausch} for inclusion (\textit{in silico}) dataset, Experiment 2, 0-30\degree.}
\begin{tabular}{ | c | c | c | c | c| c| c|}
\hline
\multirow{2}{*}{\textbf{Metric}}  &  \multicolumn{2}{c|}{\textbf{ScatGAN1}}  & \multicolumn{2}{c|}{\textbf{ScatRec1}} & \multicolumn{2}{c|}{\textbf{ScatRec7$\pm30\degree$}} \\
\cline{2-7}
& mean & max & mean &  max & mean & max \\
\cline{1-7}
MII [\%] &  3.9 & 12.5 &  34.2 & 65.0 & $\approx$2.3 & - \\
\hline
CNR [\%] &  7.3  & 11.5 &  15.1 & 40.8 & $\approx$2   & $\approx$5  \\
\hline
SNR [\%] &  2.2 & 8.1 & $\approx$8   & 15.2 & $\approx$2   & $\approx$6 \\
\hline
$\chi^2$ [$\times10^{-3}$] & 1.85 & 2.3 & $\approx$4  &  $\approx$7 & $\approx$1.2 & $\approx$2\\
\hline
Runtime [s] & 0.3 & - & 1380 & - & 9900 & - \\
\hline
\end{tabular}
\label{tab:scatrec1_comp}
\end{table}

\vspace{1ex}\noindent{\bf Comparison with Related Work.}
We quantitatively compare our results with that of~\cite{mattausch}, which performed a similar set of experiments.
Their optimization-based technique is called ``ScatRec1'' when reconstructing scatter maps from a \textit{single} B-mode image, and "ScatRec7$\pm30^{\circ}$" when reconstructing them from 7 images taken at regular beam-steering angles $\{-30^{\circ}, -25^{\circ}, ..., 30^{\circ}\}$.
Comparing the methods with single image input (ScatGAN1,ScatRec1), our pipeline is substantially better for all metrics and is $4600\times$ faster, see Table~\ref{tab:scatrec1_comp}. The speed gain mainly comes from the fact that we use a prelearned function to approximate the deconvolution, instead of solving a large system of equations for each frame. Both of the ScatRec methods face difficulties in generalizing to new beam-forming angles, resulting in high maximum error. In our method, new beam incidence angles have little impact on the performance. 
ScatGAN1 performs comparably to ScatRec7, with normalized errors mostly within a few percentage points from each other except CNR. Differences can be attributed partially to the randomness introduced during the dataset generation process, and the limited size of our training set. Our method has the advantage of not requiring the acquisition of 7 views per tissue region, and being $33000\times$ faster. This is especially useful when reconstructing large scatter maps of 3D B-mode volumes.

\section{Conclusion}
We have devised a learning-based pipeline for solving the inverse problem of US image simulation, which delivers a US tissue scatter map given an input B-mode image. Our method has shown to be relatively robust to changes in viewing parameters and US probe settings such as transducer rotation relative to the tissue region of interest. It performs well on synthetic data, and generalizes to \textit{in vivo} data with relative ease, while remaining orders of magnitude faster at inference time than the state-of-the-art optimization-based alternative~\cite{mattausch}.
Our fast runtime would make an extension to 3D feasible, which will then enable the determination of realistic input scatter maps for ray-based training simulations~\cite{gpu-raytracing}.

\bibliographystyle{IEEEbib}
\bibliography{main}
\end{document}